\documentclass[aps,pra,reprint,onecolumn,notitlepage,eqsecnum,floatfix,nofootinbib,showkeys]{revtex4-2}

\usepackage{amsmath}
\usepackage{color}
\usepackage{graphicx}
\usepackage{hyperref}
\usepackage{multirow}

\newcommand{\bq}{\begin{eqnarray}}
\newcommand{\eq}{\end{eqnarray}}
\newcommand{\bqn}{\begin{eqnarray*}}
\newcommand{\eqn}{\end{eqnarray*}}
\newcommand{\bqs}{\begin{subequations}}
\newcommand{\eqs}{\end{subequations}}
\newcommand{\bw}{\begin{widetext}}
\newcommand{\ew}{\end{widetext}}

\newcommand{\kk}{{\bf k}}
\newcommand{\rr}{{\bf r}}
\newcommand{\pp}{{\bf p}}

\newcommand{\calh}{{\cal H}}
\newcommand{\calv}{{\cal V}}
\newcommand{\calk}{{\cal K}}

\begin{document}
\title{The structure of a fluid in the canonical and in the grand canonical ensembles}

\author{Riccardo Fantoni}
\email{riccardo.fantoni@scuola.istruzione.it}
\affiliation{Universit\`a di Trieste, Dipartimento di Fisica, strada
  Costiera 11, 34151 Grignano (Trieste), Italy}

\date{\today}

\begin{abstract}
The ensemble inequivalence in a (unscreened) Coulomb liquid has been known since the years 1960s. 
The structure described by the canonical ensemble and the grand canonical ensemble are not 
equivalent. We show that this fact also affects many long range fluid. 
\end{abstract}


\maketitle
\section{Introduction}
\label{sec:intro}
Systems in $d$ dimensions with a pairwise interaction potential which decays at large distances 
as $v(r)\simeq b/r^{d+\sigma}$ as $r\to\infty$ with $-d < \sigma\leq 0$, are referred to as 
nonintegrable, or systems with long range interactions. Such systems have an ill defined 
thermodynamic limit \cite{Ruelle1963}. This may be correctly restored by applying the Kac 
prescription \cite{Kac1963}, within which the potential is rescaled by an appropriate factor 
which vanishes as the range of the interaction diverges. Or, for Coulomb systems,
\footnote{These has been solved exactly analytically in one dimension \cite{Edwards62} and at a 
special value of the coupling constant in two dimensions \cite{Jancovici81} on various surfaces: 
The plane \cite{Jancovici81}, the cylinder \cite{Choquard81}, the sphere \cite{Caillol81}, the 
pseudosphere \cite{Fantoni03jsp}, the Flamm's paraboloid \cite{Fantoni08jsp} $\ldots$.} 
\cite{Edwards62,Jancovici81,Martin1988} this problem is solved by making the liquid globally 
neutral. 
\footnote{In a one component plasma one uses a uniform neutralizing background of opposite 
charge than the total charge of the particles and in a two component plasma the two components 
will have opposite charges so as to result in a net zero total charge of the whole system.}
However, even in these cases, the energy remains non additive, i.e. the system cannot 
be divided into independent macroscopic parts, as is usually the case for short range 
interactions. Ensemble inequivalence, i.e. the possibility of observing different thermodynamic 
or structural properties depending on the statistical ensemble which describes the system, is 
one of the hallmarks of long range physics, which has been demonstrated in numerous classical 
systems \cite{Barre2001,Latella2015,Latella2017,Campa2020,Campa2022}.

In introductory books of statistical mechanics \cite{Fetter} we find that a statistical system 
can be described equally well by the four statistical ensembles: The microcanonical ensemble in 
which nothing fluctuates and it is completely constrained, the canonical ensemble where only the 
entropy $S$ fluctuates, the grand canonical ensemble where $S,N$ fluctuate, $N$ being 
the number of particles, and the isothermal isobaric ensemble where $S,V$ fluctuate, $V$ being 
the volume of the system. In a strategy of reduction of the extensive variables $(S,V,N)$ in 
favor of the intensive ones (temperature $T$, pressure $P$, chemical potential 
$\mu$) which gives rise to the thermodynamic potentials in the various ensembles, starting from 
the internal energy $E$ of the microcanonical, it is also natural to consider the completely 
unconstrained ensemble \cite{Latella2015} as the one where all three of the extensive variables 
$S,V,N$ fluctuates. Clearly, for additive systems the thermodynamic potentials are linear 
homogeneous functions of the extensive variables so that the thermodynamic potential of the 
unconstrained ensemble only makes sense for nonadditive systems such as the long range ones 
\cite{Latella2015}.

The thermodynamic properties of a fluid can be extracted by its (static) structure, i.e. the 
knowledge of its many body correlation functions \cite{Hansen-McDonald} and not vice versa.   
In this brief work we will show how the structure of long range many body systems displays an
inequivalence between its canonical and grand canonical descriptions. This fact, which affects 
computer experiments on these kind of models as well, should be stressed clearly because it is
often overlooked. Therefore, when dealing with long range systems, one should always specify 
which kind of ensemble he is using to extract the properties of the fluid he is studying.   
  
\section{Ensemble inequivalence}
Given a fluid of $N$ particles of mass $m$ in a volume $V$ with a density 
$\rho=N/V$ at an absolute temperature $T$ in the canonical ensemble, the $n$-particle density of 
the many body system is defined as \cite{Hansen-McDonald}
\bq
\rho^{(n)}_N(\rr^n)&=&\frac{N!}{(N-n)!}\frac{\int\int\exp[-\beta\calh_N(\rr^N,\pp^N)]\,d\rr^{N-n}d\pp^N}{Q_N(V,T)}\\
&=&\frac{N!}{(N-n)!}\frac{\int\exp[-\beta\calv_N(\rr^N)]\,d\rr^{N-n}}{Z_N(V,T)},
\eq
where $\rr^n=(\rr_1,\ldots,\rr_n)$ and $\pp^n=(\pp_1,\ldots,\pp_n)$ are respectively the 
coordinates and momenta of the $n$ particles, we denote with $d\rr^n=d\rr_1d\rr_2\cdots d\rr_n$,
$\beta=1/k_BT$, $\calh_N(\rr^N,\pp^N)=\calk_N(\pp^N)+\calv_N(\rr^N)$ is the Hamiltonian of 
the many body $N$ particle fluid, $Q_N=\int\int\exp[-\beta\calh_N(\rr^N,\pp^N)]\,d\rr^Nd\pp^N$
is the canonical partition function, 
and $Z_N=\int\exp[-\beta\calv_N(\rr^N)]\,d\rr^N$. So that the $\rho^{(n)}_N$ are normalized as 
$\int\rho^{(n)}_N(\rr)\,d\rr^n=N!/(N-n)!$.

The $n$ particles distribution functions are
\bq
g^{(n)}_N(\rr^n)=\rho^{(n)}_N(\rr^n)/\prod_{i=1}^n\rho^{(1)}_N(\rr_i).
\eq

From the normalization condition for $\rho^{(n)}_N$, it follows immediately that if the system 
is homogeneous, i.e. the $\rho_N^{(1)}$ are constant and equal to $\rho=N/V$,
\bq \nonumber
S_N^{c}({\bf 0})&=&1+\rho\int[g^{(2)}_N(\rr)-1]\,d\rr\\ \nonumber
&=&1+\rho\int\left[\frac{\rho^{(2)}_N(r)}{\rho^2}-1\right]\,d\rr\\ \nonumber
&=&1+\int\frac{\rho^{(2)}_N(r)}{\rho}\,d\rr-N\\ \label{eq:norm-c-lr}
&=&1+\frac{N(N-1)}{V\rho}-N=0.
\eq
where $\rr=\rr_1-\rr_2$ and $S_N(\kk)=1+\rho\int[g^{(2)}_N(\rr)-1]\exp(-i\kk\cdot\rr)\,d\rr$ is 
the static structure factor of the fluid, with $\kk$ the wave vector of the beam scattering with 
the fluid.

Eq. (\ref{eq:norm-c-lr}) is correct for Coulomb systems even in the thermodynamic limit where 
it is also known as the charge sum rule, the first of a hierarchy of sum rules called multipolar 
\cite{Martin1988} imposed by electrostatics and the hypothesis of exponential clustering, i.e. 
the asymptotic exponential decay of correlation functions as two groups of particles are 
largely separated. For long range non Coulombic potentials, $\sigma\neq -2$,
\footnote{Even if, instead of considering the Coulomb potential as the solution of the Poisson's 
equation, one could alternatively define it as $v(r)\propto 1/r$ in any dimension.} 
the Fourier transform of the pair potential $\tilde{v}(k)\simeq a_dbk^\sigma$ as $k\to 0$, and 
an analysis of the equilibrium equations of the Born-Green-Yvon hierarchy shows (see section 
II-B-3 of Ref. \cite{Martin1988}) that the Fourier transform of the static structure factor, in 
the thermodynamic limit, $S^c(k)=(\beta a_db)k^{-\sigma}+g(k)$ where $g(k)$ depends on the three 
points correlation functions and is such that $g(k)\to 0$ as $k\to 0$ due to clustering. When 
$\sigma\neq -2$ the first term is not analytic in $\kk$ at $\kk=0$ and this 
singularity introduces an algebraic term of order $r^{-(d-\sigma)}$ in the asymptotic, large
$r$, development of $S^c(r)=\int S^c(k)\exp(-i\kk\cdot\rr)\,d\kk/(2\pi)^d$ \cite{Lighthill}. 
Therefore among all possible long range potentials, it 
is only in the Coulomb case that a decay law of correlations faster than any inverse power is 
compatible with the structure of equilibrium equations. 

For short range fluids, on the other hand, we know \cite{Lebowitz1961} that, in the 
thermodynamic limit,
\bq \label{eq:LP}
\lim_{r\to\infty} g^{(2)}(\rr)=1-\frac{\chi_T}{\chi_T^0}\frac{1}{N}+ o(1/N),
\eq
where $r=|\rr|$ is the modulus of the relative vector $\rr$, 
$\chi_T=[\rho(\partial P/\partial\rho)_{N,T}]^{-1}$ is the isothermal compressibility of 
the fluid, and $\chi_T^0=\beta/\rho$ is the isothermal compressibility of the ideal gas. So 
that, in the thermodynamic limit, Eq. (\ref{eq:norm-c-lr}) should really be rewritten as 
\bq \label{eq:norm-c-sr}
S^{c}({\bf 0})=1+\rho\int[g^{(2)}(\rr)-1]\,d\rr=\frac{\chi_T}{\chi_T^0},
\eq
where we took care correctly of the additional constraint given by Eq. (\ref{eq:LP}).

On the other hand, in the grand canonical ensemble, where $P(N)$ is the probability that the 
fluid contains $N$ particles irrespective of their coordinates and momenta, we define instead 
\bq
\rho^{(n)}(\rr^n)&=&\sum_{N\geq n}P(N)\rho^{(n)}_N(\rr^n)\\
&=&\frac{1}{\Theta}\sum_{N\geq n}\frac{z^N}{(N-n)!}\int\exp[-\beta\calv_N(\rr^N)]\,d\rr^{N-n},
\eq
where $z=\Lambda^{-3}\exp(\beta\mu)$ is the activity, with $\mu$ the chemical potential and 
$\Lambda=\sqrt{2\pi\beta\hbar^2/m}$ the de Broglie thermal wavelength, and 
$\Theta=\exp[-\beta\Omega(\mu,V,T)]=\exp(\beta P V)=\sum_{N=0}^\infty\exp(N\beta\mu)Q_N$, with 
$\Omega$ the grand potential and $P$ the pressure. Then the $\rho^{(n)}$ are normalized as 
$\int\rho^{(n)}(\rr^n)\,d\rr^n=\langle N!/(N-n)!\rangle$, where $\langle\ldots\rangle$ denotes 
an average with respect to $P(N)=\exp(N\beta\mu)Q_N/\Theta$.

The average number of particles in the system is 
\bq
\langle N\rangle=\sum_{N=0}^\infty NP(N)=\frac{\partial\ln \Theta}{\partial\ln z},
\eq
so that
\bq
\frac{\partial\langle N\rangle}{\partial\beta\mu}=\langle N^2\rangle-\langle N\rangle^2.
\eq

It follows then 
\bq \label{eq:chiT}
0\leq\frac{\langle N^2\rangle-\langle N\rangle^2}{\langle N\rangle}=
\frac{1}{\langle N\rangle}\frac{\partial\langle N\rangle}{\partial\beta\mu},
\eq
where this intensive ratio is related to the isothermal compressibility. In fact, for an 
infinitesimal isothermal change it follows that $V dP = N d\mu$, where $P$ is the pressure.
If the change also takes place at constant volume, both $dP$ and $d\mu$ are proportional to $dN$:
$dP=(\partial P/\partial N)_{V,T}dN$ and $d\mu=(\partial\mu/\partial N)_{V,T}dN$. So that 
$N(\partial\mu/\partial N)_{V,T}=V(\partial P/\partial N)_{V,T}=(\partial P/\partial\rho)_{N,T}=1/\rho\chi_T$, with $\chi_T$ the isothermal compressibility. In the thermodynamic limit $N$ may be 
replaced by $\langle N\rangle$ so that \cite{Hansen-McDonald}
\bq
\frac{\langle N^2\rangle-\langle N\rangle^2}{\langle N\rangle}=
\frac{\rho\chi_T}{\beta}=\frac{\chi_T}{\chi_T^0},
\eq
with $\chi_T^0$ the isothermal compressibility of the ideal gas. 
 
The $n$ particles distribution functions are
\bq
g^{(n)}(\rr^n)=\rho^{(n)}(\rr^n)/\prod_{i=1}^n\rho^{(1)}(\rr_i).
\eq

From the normalization condition for $\rho^{(n)}$ and the thermodynamic condition of Eq. 
(\ref{eq:chiT}), it follows immediately that, if the system is homogeneous, i.e. 
$\rho^{(1)}$ is constant and equal to $\rho=\langle N\rangle/V$,
\bq \nonumber
S^{gc}({\bf 0})&=&1+\rho\int[g^{(2)}(\rr)-1]\,d\rr\\ \nonumber
&=&1+\rho\int\left[\frac{\rho^{(2)}(r)}{\rho^2}-1\right]\,d\rr\\ \nonumber
&=&1+\frac{\langle N(N-1)\rangle}{V\rho}-\langle N\rangle\\ \label{eq:norm-gc}
&=&\frac{\langle N^2\rangle}{\langle N\rangle}-\langle N\rangle=\frac{\chi_T}{\chi_T^0}.
\eq

Comparing Eq. (\ref{eq:norm-gc}) with Eqs. (\ref{eq:norm-c-sr}) and (\ref{eq:norm-c-lr}) we then 
see that the structure predicted by the canonical ensemble agrees with the one predicted by the 
grand canonical ensemble for short range fluids. For long range (unscreened) Coulomb fluids they do 
not agree giving rise to ensemble inequivalence
\footnote{Here we are thinking of unscreened charged liquids, i.e. liquids whose constituents 
particles are (not completely screened) charges. Not liquids like water or even 
like ionic screened two component liquids where one should rather look at the Bhatia-Thornton 
structure factors as done in Refs. \cite{Fantoni13f,Fantoni13e,Aqua2004}. In particular the 
work of Aqua and Fisher \cite{Aqua2004} suggests to look for ensemble inequivalence in ionic liquids 
near the critical point when there is an {\sl ionic asymmetry} which couples charge
and density fluctuations in a direct manner: The charge correlation length then diverges 
precisely as the density correlation length.}. For long range non 
Coulombic systems one should study case by case. 

\section{Conclusions}

We showed that it has long been known about the ensemble inequivalence in Coulomb liquids which 
manifests itself for example comparing the structure of the liquid in the canonical and grand 
canonical ensembles. For example we can compare the canonical path integral Monte Carlo simulation 
of the Fermi one component plasma (the Jellium) of Ref. \cite{Brown2013} where we clearly see from 
Fig. 2 that $S(0)=0$ with the grand canonical (worm algorithm) path integral Monte Carlo simulation 
of the Bose one component plasma of Ref. \cite{Zhang2023} where $S(0)\neq 0$. Recently there has 
been a revival of interest in this fact also for what concerns the inequivalence between the micro 
canonical and canonical ensembles of some long range fluids \cite{Barre2001} or the inequivalence 
between the unconstrained and isothermal isobaric ensembles of the modified Thirring model 
\cite{Latella2015,Latella2017,Campa2020,Campa2022}. In this short work we showed that long range 
fluids may or may not be inequivalent in the structure described by the canonical and the grand 
canonical ensembles. This should always be kept in mind when studying or simulating these kinds of 
long range statistical systems.

\acknowledgments
I thank my wife Laure Gouba for constant distant support and encouragement.

\section*{Author declarations}
\subsection*{Conflict of interest}
The author has no conflicts to disclose.

\section*{Data availability}
The data that support the findings of this study are available from the 
corresponding author upon reasonable request.
\bibliography{sk}

\end{document}